\documentclass[aps,prd,eqsecnum,
nofootinbib]{revtex4}

\usepackage{amsmath,amsfonts,amssymb,color,latexsym,theorem,mathrsfs,comment,color}
\usepackage[dvips]{hyperref}

\newcommand{\ma}[1]{\mbox{$\mathcal{#1}$}}
\newcommand{\mas}[1]{\mbox{$\mathscr{#1}$}}

\newcommand{\D}{{\rm d}}

\newcommand{\ti}{\tilde}
\newcommand{\we}{\wedge}

\begin{document}

\title{
Hairy Taub-NUT solution in gauged supergravity
}

\author{Masato Nozawa}
\email{
masato.nozawa@oit.ac.jp
}

\address{
Department of General Education, Faculty of Engineering, 
Osaka Institute of Technology, 5-16-1, Omiya, Asahi-ku, 
Osaka, Osaka 535-8585, Japan
}


\begin{abstract} 
We present and study a family of exact Taub-NUT-AdS solutions with scalar hair in four-dimensional ${\cal N}=2$ gauged supergravity.
The solutions are supported solely by a nontrivial scalar field and its self-interaction potential, with all gauge fields switched off.
We determine the conditions under which Killing horizons exist and demonstrate that the spacetime admits at most one Killing horizon, and whenever it exists it is necessarily nondegenerate. In particular, the NUT deformation can give rise to a horizon in a branch whose static limit is horizonless.
We also establish that the solutions with nontrivial scalar hair admit no Killing spinors. 
As in ordinary Taub-NUT-AdS spacetime, the global structure is afflicted by closed timelike curves. We compute the conserved mass using the covariant phase space method and formulate a restricted first law with the NUT parameter held fixed. Finally, we study the Euclidean sector, including an exceptional family specific to Euclidean signature, and determine the regularity conditions for bolts and nuts.  
\end{abstract}


\maketitle


\section{Introduction}

Asymptotically anti-de Sitter (AdS) black holes play a central role in holography, providing gravitational descriptions of thermal states and a geometric framework for investigating strongly coupled quantum systems~\cite{Kovtun:2004de,Ryu:2006bv,Hartnoll:2008vx}. Their importance also extends beyond holographic applications: they offer a setting in which to explore how matter fields and asymptotic boundary conditions shape the space of black-hole geometries. Understanding the range of solutions admitted by a given theory is therefore a fundamental problem, linking the dynamics of gravity to the physical content of its holographic description.

Despite this significance, the construction of exact asymptotically AdS solutions remains a formidable task. The intrinsic nonlinearity of the Einstein equations, compounded by their coupling to matter fields, severely limits the scope for analytic progress. In the stationary asymptotically flat systems, this difficulty can be overcome by exploiting the hidden symmetries of the dimensionally reduced field equations. Their nonlinear sigma-model description provides powerful solution-generating transformations that map known geometries into new ones~\cite{Breitenlohner:1987dg}. A cosmological constant or a nontrivial scalar potential, however, generally breaks the symmetries underlying these transformations, obstructing their straightforward extension to the asymptotically AdS setting~\cite{Klemm:2015uba}. One is then often led to an ansatz-based approach, in which the existence of an exact solution is far from guaranteed.

In gauged supergravity coupled to scalar multiplets, this obstruction is closely tied to the mechanism that supports the AdS asymptotics. The gauging generates a scalar potential whose negative critical points define AdS vacua~\cite{Trigiante:2016mnt,DallAgata:2021uvl}. The difficulty is sharpened in gauged supergravity, because the potential cannot be engineered at will but is severely constrained  by supersymmetry. Constructing a solution with a nontrivial scalar profile consequently requires reconciling the nonlinear gravitational dynamics with these restrictions on the scalar self-interactions.

The scalar potential that obstructs familiar routes to exact solutions can also open the door to a richer landscape of black holes. For minimally coupled scalar fields, familiar no-hair results in asymptotically flat spacetime rely on specific assumptions about the scalar potential and the asymptotic behavior of the field~\cite{Bekenstein:1971hc}. An appropriate potential together with AdS boundary conditions can evade these restrictions, allowing nontrivial scalar hair to coexist with a regular horizon even in the absence of electromagnetic fields~\cite{Hertog:2004dr,Anabalon:2012sn,Anabalon:2012ta,Anabalon:2013eaa,Feng:2013tza,Faedo:2015jqa,Anabalon:2020pez}. The question is therefore not merely whether scalar-haired black holes can exist, but how diverse their geometries can be once the theory and its boundary conditions are fixed. Despite substantial progress, a systematic understanding of this diversity in gauged supergravity remains incomplete, particularly beyond the supersymmetric sector.

The construction of exact static scalar-haired black holes in gauged supergravity naturally motivates the search for rotating generalizations.
Obtaining exact Kerr-like solutions in the presence of a nontrivial scalar potential, however, is technically demanding.
An alternative route beyond staticity is to introduce Newman--Unti--Tamburino (NUT) charge~\cite{Taub:1950ez,Newman:1963yy}, which allows a cohomogeneity-one description. Despite the distinction from ordinary angular momentum, NUT charge plays the role of a
gravitomagnetic charge \cite{Dowker:1974znr,Argurio:2008zt,Galtsov:2026qlf} and enters through a nontrivial fibration of the time direction over the angular sector, modifying both the asymptotic geometry and the global structure of spacetime. Its inclusion therefore does more than enlarge the parameter space of a scalar-haired solution: it raises the question of how scalar dynamics interact with a geometric deformation that affects the causal structure and the definition of conserved quantities. NUT-charged configurations thus provide a useful setting in which to examine this interplay, while requiring a careful distinction between local regularity
and global consistency.

In this paper, we present a family of  exact solutions with NUT charge and scalar hair in four-dimensional ${\cal N}=2$ gauged supergravity. 
NUT-charged configurations with scalar hair have also been explored in other theories of gravity~\cite{Brihaye:2018bgc,Barrientos:2022avi}.
The present solutions are asymptotically locally AdS and belong to an Einstein-scalar sector in which no electromagnetic fields are turned on. Their nontrivial matter content is thus supplied entirely by the scalar field and its self-interaction potential. 
The solutions can also be recovered from the cohomogeneity-two solution in the more general Einstein-scalar theory of \cite{Anabalon:2012ta}, 
 by specializing the theory parameters and taking a nontrivial cohomogeneity-one scaling limit. To the best of our knowledge, however, their explicit cohomogeneity-one form has not previously been presented. Here we provide an independent derivation of this sector and investigate
its physical properties within the specified supergravity theory. In particular, we determine the conditions for the existence of regular 
Killing horizons, examine the geometric and causal structure, and compute the conserved quantities.
These solutions furnish an exact testbed for investigating the interplay between scalar hair, NUT charge, and AdS asymptotics within a theory whose
interactions are constrained by supersymmetry. 

We also inquire whether a nontrivial scalar profile can be retained in smooth Euclidean geometries with a NUT fibration. 
In addition to the direct analytic continuations of the Lorentzian solutions, we obtain an exceptional NUT-charged family through a
scaling limit intrinsic to the Euclidean signature. For special values of parameters 
admitting higher-dimensional supergravity embeddings, we determine the conditions under which these geometries close off smoothly at a bolt or a nut. 
Compatibility between regularity at these fixed-point sets and the periodic identification required to remove the Dirac-Misner strings
constrains both the continuous parameters and the topology of the resulting geometries.

The remainder of this paper is organized as follows. 
In the next section, we introduce the Lagrangian, present the solutions, and clarify their relation to previously known families.
In section~\ref{sec:phys}, we examine their physical properties, including the singularity structure, Killing horizons, the absence of Killing
spinors, Dirac-Misner strings, geodesic motion, and asymptotic behavior. 
We compute the conserved mass using the covariant phase space method and discuss black-hole thermodynamics with the NUT parameter held fixed.
Section~\ref{sec:Euc} examines the Euclidean continuations of the solutions. We present an additional family of NUT-charged geometries specific
to Euclidean signature, together with an analysis of their bolt and nut structures.
Section~\ref{sec:conclusion} summarizes our results with some future outlooks.

\section{Taub-NUT solution in gauged supergravity}
\label{sec:TN}

\subsection{${\cal N}=2 $ gauged supergravity}

In this paper, we consider the gravitational solutions in Einstein-scalar gravity described by the Lagrangian
\begin{align}
\label{Lag}
\ma L= R -\frac 12 (\nabla \phi)^2 -2V(\phi) \,,  
\end{align}
where $V(\phi)$ is a scalar potential written in terms of a real superpotential $W=W(\phi)$ as
\begin{align}
\label{pot}
V(\phi)=4 \left[4W'(\phi)^2 -3W(\phi)^2 \right] \,.
\end{align}
Einstein's equations and the scalar equation read
\begin{align}
\label{}
R_{\mu\nu}=\frac 12 \nabla_\mu \phi \nabla_\nu \phi +V(\phi)g_{\mu\nu} \,, \qquad 
\nabla^2 \phi-2 V'(\phi)=0 \,. 
\end{align}

Specifically, we consider the following form of the superpotential \cite{Faedo:2015jqa,Lu:2013eoa,Nozawa:2022upa}
\begin{align}
\label{W}
W(\phi)=\frac{g}{2(1+\alpha^2)}\left(e^{\frac \alpha 2\phi}+\alpha^2 e^{-\frac{\phi}{2\alpha}}\right)\,, 
\end{align}
where $\alpha$ and $g$ are constants. 
This theory is obtained by a truncation of ${\ma N}=2$ Fayet-Iliopoulos gauged supergravity with the prepotential \cite{Faedo:2015jqa}
\begin{align}
\label{prepotential}
\ma F(\ma X)=-\frac i 4 (\ma X^0)^{\frac{2}{1+\alpha^2}}(\ma X^1)^{\frac{2\alpha^2}{1+\alpha^2}} \,.
\end{align}
For the truncation, one sets the two gauge fields and the axion field to zero.  
In the context of extended supergravity, the constant $g$ plays multiple roles: it sets the fermion mass scale, serves as the gauge coupling constant, and determines the overall scale of the scalar potential~\cite{Trigiante:2016mnt,DallAgata:2021uvl}.

The Lagrangian is invariant under
\begin{align}
\label{sym1}
\alpha \to -\alpha \,, \qquad \phi \to -\phi \,, 
\end{align}
and 
\begin{align}
\label{sym2}
\alpha \to -\frac 1\alpha \,.
\end{align}
The symmetry (\ref{sym1}) enables us to focus on the $\alpha >0$ case without loss of generality. 
Although  one can further restrict to $\alpha\ge 1$ or $0<\alpha \le 1$ using the second symmetry (\ref{sym2}),
we will leave this freedom unfixed for convenience.  
In the special cases $\alpha =1$ and $\alpha=\sqrt 3 $ (or $\alpha=1/\sqrt 3$), 
this theory can be obtained by the truncation of ${\cal N}=4$ and ${\cal N}=8$ supergravities respectively, 
 and can be embedded into $D=11$ dimensions \cite{Duff:1999gh,Cvetic:1999au}.

The origin $\phi=0$ of the scalar field corresponds to a critical point of the scalar potential and the superpotential.\footnote{
The potential $V$ may admit another AdS vacuum at $\phi=2\alpha (1+\alpha^2)^{-1} \ln [(3\alpha^2-1)/(\alpha^2-3)]$
for $\alpha^2<1/3$ or $\alpha^2>3$. However, this vacuum is not our present concern. }
At the origin, we have
\begin{align}
\label{masseig}
V(0)=-3g^2 \,, \qquad m_\phi^2 \equiv 2V''(0)=-2g^2 \,. 
\end{align}
Thus, $g^{-1}$ corresponds to the AdS radius at the origin.
 It is notable that the mass parameter $m_\phi$ lies in a characteristic window
 \begin{align}
\label{}
m_{\rm BF}^2<m_\phi^2<m^2_{\rm BF}+g^{2} \,,  
\end{align}
where $m^2_{\rm BF}=-9g^2/4$ corresponds to the Breitenlohner-Freedman bound
\cite{Breitenlohner:1982jf,Ishibashi:2004wx}. In this mass range, the scalar field 
$\phi$ may obey Robin-type mixed boundary conditions at infinity and two asymptotic scalar modes
are both normalizable. 

In this paper, we wish to construct a gravitational solution with NUT charge which asymptotically tends to the AdS vacuum at $\phi=0$. 
Since the scalar potential obstructs the conventional coset-based solution-generating methods, we are forced to employ an ansatz-based construction. Even within this restricted framework, however, the existence of an exact asymptotically AdS solution with a nontrivial scalar profile is by no means guaranteed. The fact that such a solution can nevertheless be obtained is therefore significant, providing a rare exact testbed for exploring the interplay between NUT charge, scalar hair, and asymptotically AdS boundary conditions.

\subsection{Taub-NUT solution with scalar hair in AdS}

To construct the new Taub-NUT-AdS solution with scalar hair, we employ the approach 
initiated in \cite{Nozawa:2020gzz}. The procedure introduced in \cite{Nozawa:2020gzz} begins by expressing the asymptotically flat seed solution in isotropic coordinates and incorporating it into a McVittie ansatz with asymptotically de Sitter behavior. The scalar potential is then fixed, a posteriori,  by requiring the field equations to be satisfied. After transforming the solution back to static coordinates, one obtains the associated asymptotically AdS geometry through an analytic continuation of the Hubble parameter. Compared to the asymptotically flat case, the resulting static metric is dressed with a ``blackening factor'', in which the presence of the scalar potential is encoded.  Remarkably, the resulting scalar potential is expressed in terms of a superpotential like (\ref{pot}), although the explicit form of $W$ depends on the choice of the seed solution. Moreover, changing the sign of the scalar field and solving Einstein's equations again with the same scalar potential while leaving the blackening factor unspecified, one can find another branch of asymptotically AdS solutions. 
Recently, this prescription was extended to the stationary case of wormholes with a NUT parameter~\cite{NHN}, by taking the 
Ellis-Bronnikov wormhole \cite{Ellis:1973yv,Bronnikov:1973fh,Martinez:2020hjm} as a seed solution. Note that in this stationary case the McVittie ansatz does not work, but the blackening prescription does.

Adopting the recently constructed NUT-charged solution with scalar hair \cite{Bakherad:2026crn} as a seed, 
this recipe successfully generates a new solution of the present system (\ref{Lag})-(\ref{W}):
\begin{align}
\label{metric}
\D s^2=&\,-f(r) \Delta_\sigma (r) \left(\D t+2 n \gamma \cos\theta \D \varphi \right)^2
+\frac 1{f(r)}\left[\frac{\D r^2}{\Delta_\sigma (r)}+r^2 f_0(r) \left(\D \theta ^2+\sin^2\theta \D \varphi^2\right)\right]\,, 
\\
\label{scalar}
\phi =&\, \sigma \frac{2\alpha}{1+\alpha^2} \ln f_0(r) \,, 
\end{align}
where  $\sigma =\pm 1$ denotes the branch, and 
\begin{align}
\label{f0}
f_0(r)=&\, 1-\frac{2m_0}{r} \,, 
\\
\label{f}
f(r)=&\, \frac{2m_0 f_0(r)^{\gamma}}{(m_0-m)f_0(r)^{2\gamma}+m_0+m}\,, 
\\
\Delta_\sigma(r)=&\, 1+g^2 \left[\frac{m_0+\sigma m}{2m_0}r^2 f_0(r)^{1-2\sigma \gamma}+\frac{m_0-\sigma m}{2m_0}
\left\{[r-(1+2\sigma \gamma )m_0]^2+m_0^2(4\gamma^2-1)\right\}
\right] \,.\label{Delta}
\end{align}
Here, the parameter $m_0$ is given by
\begin{align}
\label{m0}
m_0=\sqrt{m^2+n^2} \,. 
\end{align}
For convenience, we have introduced 
\begin{align}
\label{gamma}
\gamma \equiv \frac{1-\alpha^2}{1+\alpha^2}\,, \qquad 
-1\le \gamma \le 1 \,. 
\end{align}
In particular, $\alpha=1$ corresponds to $\gamma=0$, while 
$\alpha=\sqrt 3$ and $1/\sqrt 3$ correspond to $\gamma=-1/2$ and $\gamma=1/2$, respectively. 
These cases ultimately admit embeddings into eleven-dimensional supergravity. 

The angular coordinates $\theta$ and $\varphi$ parametrize the standard two-sphere $S^2$. 
The function $\Delta_\sigma$ plays the role of a blackening factor, which is responsible for the 
existence of horizons.

\subsection{Comparison with the literature}

Before turning to a detailed analysis of the physical properties of the solution,
it is instructive to clarify its relation to those obtained in the literature.

\subsubsection{Taub-NUT-AdS limit $|\gamma|\to 1$}

Setting $|\gamma|=1$ (i.e., $\alpha=0$ or $\alpha=\pm \infty$), the scalar field given in (\ref{scalar}) becomes trivial. 
One can then verify that the metric reduces to the standard Taub-NUT-AdS spacetime~\cite{Brill,Griffiths:2009dfa}
\begin{align}
\label{TNAdS}
\D s^2=-f_\Lambda (\tilde r)(\D t+2 n \cos\theta \D \varphi)^2+\frac{\D\tilde  r^2}{f_\Lambda (\tilde r)}+(\tilde r^2+n^2)
(\D \theta^2+\sin^2\theta \D \varphi^2)\,, 
\end{align}
with
\begin{align}
\label{}
 f_\Lambda (\tilde r)=\frac{\tilde r^2-2\tilde m \tilde r-n^2}{\tilde r^2+n^2}+g^2\frac{\tilde r^4+6n^2\tilde r^2-3n^4}{\tilde r^2+n^2}\,,
\end{align}
by the coordinate transformation
\begin{align}
\label{tirm}
\tilde r=r-\sqrt{m^2+n^2}+{\rm sgn}(\gamma)m \,, \qquad 
\tilde m\equiv  {\rm sgn}(\gamma)\left\{m+2g^2\left[m(m^2+3n^2)-\sigma(m^2+n^2)^{3/2}\right]\right\}\,.
\end{align}
The spacetime (\ref{TNAdS}) satisfies $R_{\mu\nu}=-3g^2 g_{\mu\nu}$ and is free from
scalar polynomial curvature singularities.
Notice also that the branch label $\sigma$ survives only
through the redefinition of the mass parameter $\tilde m$ and therefore does
not characterize an independent branch of the Taub-NUT-AdS geometry.

\subsubsection{ALF limit $g\to 0$}

Taking the $g\to 0$ limit, the scalar potential vanishes and the solution reduces to
\begin{align}
\label{metric0}
\D s^2=&\,-f(r) \left(\D t+2 n \gamma \cos\theta \D \varphi \right)^2
+\frac 1{f(r)}\left[\D r^2+r^2 f_0(r) \left(\D \theta ^2+\sin^2\theta \D \varphi^2\right)\right]\,, 
\\
\label{scalar0}
\phi =&\, \pm \frac{2\alpha}{1+\alpha^2} \ln f_0(r) \,, 
\end{align}
where $f_0$ and $f$ are still given by (\ref{f0})  and (\ref{f}). 
This solution is asymptotically locally flat (ALF) and was constructed recently in \cite{Bakherad:2026crn} by means of an Ehlers-type transformation. The deformation induced by the scalar potential is encoded entirely in the function $\Delta_\sigma$, while the remaining functions $f(r)$ and $f_0(r)$ retain precisely their ALF forms. Contrary to the $g\ne 0$ case, the ALF solution always describes a naked singularity. 

It is worth emphasizing that the role of the parameter $\gamma$ changes in this limit. For $g\neq 0$, $\gamma$ enters the scalar potential [see (\ref{pot}) and (\ref{gamma})] and thus characterizes the theory itself, whereas $\gamma $ decouples from the Lagrangian in the ALF limit since the potential vanishes.  $\gamma$ is therefore demoted to a mere parameter labeling the solution. Likewise, the distinction between the two branches disappears in the ALF limit. Since the massless scalar theory is invariant under $\phi\to-\phi$, the label $\sigma=\pm1$ merely corresponds to this discrete symmetry and does not distinguish physically inequivalent configurations. In particular, the two branches yield the same metric.

\subsubsection{Static limit $n\to 0$}
\label{sec:static}

Taking the limit $n\to 0$, our solution becomes static and reduces to 
\begin{align}
\label{}
\D s^2=&\,-f_0(r)^\gamma \Delta_\sigma (r) \D t^2
+\frac 1{f_0(r)^\gamma}\left[\frac{\D r^2}{\Delta_\sigma (r)}+r^2 f_0(r) \left(\D \theta ^2+\sin^2\theta \D \varphi^2\right)\right]\,, 
\end{align}
where 
the scalar field is still given by (\ref{scalar}) with $f_0(r)=1-2m/r$, and 
\begin{align}
\label{}
\Delta_+(r)=1+g^2r^2f_0(r)^{1-2\gamma} \,, \qquad 
\Delta_-(r)=1+g^2 \left\{\left[r-m(1-2\gamma)\right]^2+m^2(4\gamma^2-1)\right\}\,. 
\end{align}
Here, we have assumed $m>0$ for definiteness and to conform with the branch convention used in the literature.
This restriction will be relaxed below when a single branch is considered. 

In contrast to the ALF and Taub-NUT-AdS limits discussed above, the two branches of solutions are physically inequivalent. 
The plus-branch  solution ($\sigma=+1$) recovers
the one in \cite{Anabalon:2012ta,Feng:2013tza}. This solution is horizonless and describes a nakedly singular spacetime in AdS, 
since $\Delta_+(r)$ is positive definite. The minus-branch solution ($\sigma=-1$) reduces 
to the AdS hairy black hole solution in \cite{Faedo:2015jqa} with a regular event horizon $r_{\rm h}$ at $\Delta_-(r_{\rm h})=0$, provided
\begin{align}
\label{statichorizon}
0<|\gamma|<\frac 12\,, \qquad  2 mg\gamma < -\sqrt{\frac{|\gamma|}{1-2|\gamma|}}\,.
\end{align}
These conditions are obtained by requiring $r_{\rm h}$ to lie outside the outermost curvature singularity 
$r=\max(0,2m)$.

Unlike in the Taub-NUT-AdS and ALF limits, the branch label $\sigma=\pm1$ remains physically meaningful and distinguishes qualitatively different solutions within the same underlying theory.\footnote{
The origin of these two branches for static solutions can be understood from the corresponding C-metric solutions of the same theory. As shown in \cite{Nozawa:2022upa} (see also \cite{Nozawa:2023aep}), the two families are related by a flipping transformation characteristic of the C-metric coordinates, which exchanges the radial and angular sectors accompanied by a double Wick rotation. This transformation ceases to be available in the zero-acceleration limit, thereby giving rise to two distinct static solutions within the same theory.
} As we will see below, this distinction persists also in the stationary NUT-charged solutions.

\subsubsection{Cohomogeneity-one limits of the solutions in
\cite{Anabalon:2012ta}}

The present solution can be obtained as a cohomogeneity-one scaling
limit of the more general stationary and axisymmetric Einstein-scalar
family constructed in \cite{Anabalon:2012ta}.
To exhibit this relation, we consider its minimally coupled sector.
In the scalar normalization used in this paper, the metric and scalar field are
\begin{align}
\label{Anabalonmetric}
\D s^2={}&
-\frac{C}{6}
\frac{p^{\nu-1}q^{\nu-1}}{(C_3p^\nu+q^\nu)^2}\Biggl[(1+p^2q^2)\left(\frac{\D p^2}{X(p)}+\frac{\D q^2}{Y(q)}\right)
\notag\\
&\qquad
-\frac{Y(q)}{1+p^2q^2}(p^2\D\tau+\D\varsigma)^2+\frac{X(p)}{1+p^2q^2}(\D\tau-q^2\D\varsigma)^2\Biggr]\,,
\\
\label{Anabalonscalar}
\varphi={}&
\eta\sqrt{\nu^2-1}\,\ln\left(h\frac{q}{p}\right)\,,\qquad \eta=\pm1\,,
\end{align}
where
\begin{align}
\label{AnabalonX}
X(p)={}& C_0+C_2p^2+C_4p^4+C_1p^{2-\nu}+C_3B_3p^{\nu+2}\,,
\\
\label{AnabalonY}
Y(q)={}&C_4-C_2q^2+C_0q^4+C_3C_1q^{2-\nu}+B_3q^{\nu+2}\,.
\end{align}
The scalar potential is given by Eq.~(25) of \cite{Anabalon:2012ta}, 
which in our normalization takes the form
\begin{align}
\label{AnabalonPotential}
V_\eta(\phi)
=&\frac{3}{C}\Bigg[
 C_0\frac{(\nu-1)(\nu-2)}{h^{\nu+1}} e^{\frac{(\nu+1)\eta\phi}{\sqrt{\nu^2-1}}}
+C_4\frac{(\nu+1)(\nu+2)}{h^{\nu-1}} e^{\frac{(\nu-1)\eta\phi}{\sqrt{\nu^2-1}}}
-4C_3C_0\frac{(\nu^2-1)}{h} e^{\frac{\eta\phi}{\sqrt{\nu^2-1}}}
 \notag\\
&\quad
 +C_3^2C_0\frac{(\nu+1)(\nu+2)}{h^{1-\nu}} e^{-\frac{(\nu-1)\eta\phi}{\sqrt{\nu^2-1}}}
 +C_3^2C_4\frac{(\nu-1)(\nu-2)}{h^{-1-\nu}} e^{-\frac{(\nu+1)\eta\phi}{\sqrt{\nu^2-1}}}
 -4C_3C_4\frac{(\nu^2-1)}{h^{-1}} e^{-\frac{\eta\phi}{\sqrt{\nu^2-1}}}
 \Bigg]\,.
\end{align}
Here, $\eta=+1$ corresponds to the scalar branch displayed explicitly
in \cite{Anabalon:2012ta}, whereas $\eta=-1$ is obtained by reversing
the sign of the scalar field. 
It is manifest from (\ref{AnabalonPotential}) that
$\nu$, $C$, $C_0$, $C_3$, $C_4$, and $h$ enter the scalar potential and thus 
correspond to the theory parameters, whereas $C_1$, $C_2$, and $B_3$ are integration constants. 
For definiteness, we work in a region with $p,q,h>0$ and take $\nu^2>1$.

The reduction proceeds in two steps. We first specialize the theory parameters so that the scalar potential agrees with (\ref{pot}), and then take a scaling limit around a double root of the angular structure function, whereby the angular sector becomes a round two-sphere while a finite NUT fibration survives.

The generic potential is more general than (\ref{pot}), and its parameters must first be specialized to select the
theory considered here. It can be verified that the potential (\ref{AnabalonPotential}) reduces to (\ref{pot}), 
provided 
\begin{align}
\label{Anabalontheory}
C=-6\,,
\quad C_3=-1\,,
\quad h=1\,,
\quad \nu=\frac{1}{\gamma}\,,
\quad \eta= \sigma\,,
\quad 
C_0=\frac{1-\sigma}{2}g^2\gamma^2\,,
\quad
C_4=\frac{1+\sigma}{2}g^2\gamma^2\,.
\end{align}
For simplicity of the argument, we specialize to the $\gamma>0$ case below. 
The case $\gamma<0$ can be treated analogously and leads to the same local family.

We now argue that the cohomogeneity-two solution (\ref{Anabalonmetric}) 
reduces to our cohomogeneity-one solution. 
To take this limit, introduce constants $p_0>0$, $K>0$, and a small parameter $\varepsilon$, and
parametrize the remaining integration constants as
\begin{align}
\label{AnabalonC1}
C_1={}&
C_1^{(0)}+Kp_0^{\nu-2}\varepsilon^2\,, \\
\label{AnabalonC10}
C_1^{(0)}={}&
\frac{p_0^{\nu-2}}{\nu^2}\left[-Kp_0^2-(\nu+2)C_0+(\nu-2)C_4p_0^4 \right]\,,
\\
\label{AnabalonB3}
B_3={}&
\frac{p_0^{-\nu-2}}{\nu^2}
\left[Kp_0^2-(\nu-2)C_0+(\nu+2)C_4p_0^4 \right]\,,
\\
\label{AnabalonC2}
C_2={}&\frac{2Kp_0^2-(\nu^2-4)(C_0+C_4p_0^4)}{\nu^2p_0^2}\,.
\end{align}
We denote $X_0$ and $Y_0$ for the $\varepsilon\to 0$ limit of structure functions $X(p)$ and $Y(q)$.
The above parametrization ensures $X_0(p_0)=X_0'(p_0)=0$ and $X_0''(p_0)=-2K$. 
The limit therefore focuses on a neighborhood of a double root of the angular structure function.

We now introduce the coordinates
\begin{align}
\label{Anabalonangularscaling}
p=p_0+\varepsilon\cos\theta\,,
\qquad
\tau=\frac{\varphi}{K\varepsilon}\,,
\qquad
\varsigma=bt-\frac{p_0^2}{K\varepsilon}\varphi\,,
\end{align}
where $b\neq0$ fixes the normalization of the time coordinate.
As $\varepsilon\to0$, one has
$X(p)=K\varepsilon^2\sin^2\theta+O(\varepsilon^3)$, 
while $p^2\D\tau+\D\varsigma \to b\D t+2p_0K^{-1}\cos\theta\D\varphi$ and 
$\D \tau-q^2 \D \varsigma\to \varepsilon^{-1}K^{-1}(1+p_0^2 q^2)\D \varphi$.
Thus, the angular sector becomes a round two-sphere,
while a finite NUT fibration survives.
Since $\varepsilon$ enters only the integration constant $C_1$, this limit is taken within a fixed theory.

The limiting metric and scalar field are
\begin{align}
\label{AnabalonNUTlimit}
\D s^2={}&
\frac{p_0^{\nu-1}q^{\nu-1}}
{(q^\nu-p_0^\nu)^2}
\Biggl[
(1+p_0^2q^2)\frac{\D q^2}{Y_0(q)}
+\frac{1+p_0^2q^2}{K}
\left(
\D\theta^2+\sin^2\theta \D\varphi^2
\right)
-\frac{Y_0(q)}{1+p_0^2q^2}
\left(
b\D t+\frac{2p_0}{K}\cos\theta\D\varphi
\right)^2
\Biggr]\,,
\\
\label{Anabalonscalarlimit}
\varphi={}&
\eta\sqrt{\nu^2-1}\ln\left(\frac{q}{p_0}\right)\,,
\end{align}
where $Y_0(q)=C_4-C_2q^2+C_0q^4-C_1^{(0)}q^{2-\nu}+B_3q^{\nu+2}$. 
Finally, setting
\begin{align}
\label{Anabalonparameteridentification}
p_0^2=\frac{n}{m_0+m}\,,
\qquad
K=\frac{1}{2m_0n}\,,
\qquad
b=\frac{2m_0p_0}{\gamma}\,,\qquad 
q=p_0f_0(r)^\gamma\,, 
\end{align}
we establish that
(\ref{AnabalonNUTlimit}) coincides with (\ref{metric}).

The above computation identifies a NUT-charged, cohomogeneity-one
sector within the broader Einstein-scalar family of \cite{Anabalon:2012ta}.
While the scalar potential (\ref{AnabalonPotential}) of that family is not restricted a priori by a supergravity embedding (into electric frame), 
the parameter choices made above select precisely the gauged-supergravity model considered here.
The present solutions are therefore contained in the known general family, rather than constituting an independent family
of local solutions. The contribution of the present work lies in their independent
derivation from a NUT-charged asymptotically locally flat seed
and in the detailed analysis of the resulting two-branch sector
within this specific supergravity theory.

\subsubsection{Remarks}

A NUT-charged black hole that preserves supersymmetry was constructed in \cite{Colleoni:2012jq}. 
Up to normalization, their prepotential coincides with ours (\ref{prepotential}) with $\alpha=1$ ($\gamma=0$). In this case, however, the present solution (\ref{metric}) reduces to a static configuration with a vanishing NUT parameter. 
Hence, our solution is complementary to the solution in~\cite{Colleoni:2012jq}. 

Nevertheless, a distinct family of NUT-like solutions for $\gamma=0$ is available 
in Euclidean signature, as we will discuss in section~\ref{sec:Euc}.


\section{Physical properties}
\label{sec:phys}

In this section, we explore physical properties of the stationary solution (\ref{metric}). 
For fixed $g$ and $\gamma$, the solution is specified by 
two parameters $m$ and $n$, which we shall refer to as the mass and NUT parameters. 
The metric is invariant under
\begin{align}
\label{symsol}
m\to -m \,, \qquad n\to -n \,, \qquad \gamma \to -\gamma \,, \qquad 
\sigma \to -\sigma \,, 
\end{align}
while the scalar field changes sign $\varphi\to-\varphi$.
Since the theories related by $\gamma\to-\gamma$ are equivalent under
a scalar-field reflection, this transformation maps the full family of solutions into itself. 
Moreover, the sign change $n\to -n$ can be compensated by reversing the
direction of rotation $\varphi\to -\varphi$. Since $\sigma=\pm1$ labels
the two branches of the solution, the  transformation (\ref{symsol}) should be
understood as a symmetry exchanging the two branches rather than as a
symmetry within each branch separately. Taking both branches into account,
these symmetries allow us to restrict attention to the range\footnote{
Since $\gamma$ is a parameter specifying the theory, it would naturally be held fixed when comparing different solutions within a given theory. For the purpose of analyzing the physical properties of the full
two-branch family, however, it is considerably more convenient to use the
above branch-exchanging symmetry to fix the sign of the solution parameter
$m$ instead. Such a restriction would not be justified if only one of the
two branches were available.}
\begin{align}
\label{mnrange}
m \ge 0 \,, \qquad n \ge 0 \,. 
\end{align}

It is obvious from (\ref{m0}) that the function $f(r)$ is nonnegative
\begin{align}
\label{}
f(r) \ge 0 \,. 
\end{align}
The causal structure therefore depends sensitively on the zeros and signs of
$f_0(r)$ and $\Delta_\sigma(r)$.

\subsection{Curvature singularities and algebraic structure}

Curvature singularities can be identified from scalar polynomial curvature invariants. 
For instance, the scalar curvature is written as 
\begin{align}
\label{Ricciscalar}
R=\frac{2m_0^2(1-\gamma^2)f(r)\Delta_\sigma(r)}{r^4 f_0(r)^2}
-4g^2 f_0(r)^{-1-\sigma \gamma}\left[3-\frac{6(1+\sigma \gamma)m_0}{r}
+\frac{2m_0^2(1+3\sigma \gamma+2\gamma^2)}{r^2} 
\right] \,.
\end{align}
For a nontrivial scalar field $\gamma^2\ne 1$, 
it follows that the curvature singularities are present at 
\begin{align}
\label{sing}
r=0 \,, \qquad  r=2m_0(>0) \,. 
\end{align}
Hereafter, we shall consider the outside region $r\ge 2m_0$.

Since the remaining curvature invariants do not have particularly illuminating expressions, we do not display them here. 
Nevertheless, one can verify that they remain finite away from the loci (\ref{sing}), so that no additional scalar polynomial
curvature singularities occur. 
This behavior contrasts with the ordinary Taub-NUT-AdS metric ($\gamma^2=1$), for which
the scalar polynomial curvature invariants remain finite everywhere.

It is also worth noting that the Weyl tensor of the metric (\ref{metric}) is of
Petrov type D, as in the Taub-NUT-AdS case. Moreover, one can verify that the following null
vectors are tangent to shear-free null geodesic congruences and coincide
with the principal null directions of the Weyl tensor
\begin{align}
\label{principalnull}
 k_\pm =\frac{1}{f(r)\Delta_\sigma (r)}\frac{\partial }{\partial t}\pm \frac{\partial }{\partial r}\,. 
\end{align}
Thus, despite the existence of the nontrivial scalar field and its potential, 
the solution is categorized into the same algebraically special family 
as the customary Taub-NUT-AdS spacetime.

\subsection{Killing horizons}
\label{sec:horizon}

Since $f(r)>0$, a Killing horizon associated with the stationary Killing vector
$\partial/\partial t$ can occur only at
\begin{align}
\label{}
\Delta_\sigma(r)=0 \,.
\end{align}
At such a zero, the hypersurface is null, while the curvature invariants remain finite.
Hence, any zero of $\Delta_\sigma(r)$ in the physical region corresponds to a locally regular
Killing horizon. For $g=0$, it is obvious that no Killing horizons appear.

Let us now reveal the conditions under which the regular horizons exist for $g\ne 0$. 
It follows directly from (\ref{Delta}) that no horizon exists for $|\gamma|\ge 1/2$, 
since $\Delta_\sigma(r)$ is a sum of nonnegative terms in this case.
We therefore restrict attention to $|\gamma|<1/2$ and examine the monotonicity of $\Delta_\sigma(r)$ in the exterior region $r>2m_0$.
A direct computation yields
\begin{align}
\label{}
\Delta_\sigma'(r)=2 g^2 f(r)^{-1} f_0(r)^{-\sigma \gamma }
\left[r-m_0(1+2\sigma \gamma )\right]\,.
\end{align}
The stationary point of $\Delta_\sigma(r)$ is thus located at $r=m_0(1+2\sigma\gamma)$.
Due to $-1/2<\sigma\gamma<1/2$, this point lies inside the boundary $r=2m_0$, and hence $\Delta_\sigma(r)$ is monotonically increasing throughout $r>2m_0$. 
Since $\Delta_\sigma(r)>0$ for large $r$,  the existence of the horizon in the region $r>2m_0$ is tantamount to
\begin{align}
\label{Delta2m0}
\Delta_\sigma(2m_0)
=1+2g^2m_0(m_0-\sigma m)\sigma\gamma(2\sigma\gamma-1)<0\,,
\end{align}
where we have used $f_0(2m_0)^{1-2\sigma\gamma}=0$ for $| \gamma|<1/2$. 
Accordingly, a horizon exists if and only if 
\begin{align}
\label{condhorizon}
0<\sigma \gamma <\frac 12 \,, \qquad 
2g^2 m_0(m_0-\sigma m)\sigma \gamma (1-2\sigma \gamma)>1 \,. 
\end{align}
Under these conditions, $\Delta_\sigma(r)$ has a unique zero at 
$r=r_{\rm h}>2m_0$. Moreover, the root is necessarily simple, 
since $\Delta_\sigma(r)$ is strictly monotonic there. Thus, the solution admits no regular
extremal Killing horizon.

Notice that, owing to the restriction $m\geq 0$ adopted in (\ref{mnrange}), the static limit $n\to0$ of (\ref{condhorizon}) reproduces the $\gamma<0$ representative of the $\sigma=-1$ branch in (\ref{statichorizon}). The corresponding $\gamma>0$ solution is represented, before fixing the sign of $m$, by $m<0$ and $\sigma=+1$, and is mapped to the former by the branch-exchanging symmetry (\ref{symsol}). Thus, no static black-hole solutions are lost by the restriction (\ref{mnrange}).

It is worth emphasizing that a regular horizon may also occur in the plus
branch ($\sigma=+1$). This is in sharp contrast to the static limit discussed
in section~\ref{sec:static}, where the plus branch invariably describes a
nakedly singular spacetime. The NUT deformation can therefore qualitatively
alter the causal structure and cloak the curvature singularity behind a
regular horizon.

Finally, in the ALF limit $g=0$, the horizon function becomes trivial
$\Delta_\sigma(r)=1$, and no horizon is present. Thus, although the NUT
deformation is responsible for the emergence of a horizon in the plus
branch, a nonvanishing scalar potential is also essential for the existence of a 
 regular horizon found above.

\subsection{Nonexistence of Killing spinors}
\label{sec:Killingspinor}

Since the present system (\ref{Lag})-(\ref{W}) is derived from supergravity, 
it is interesting to explore if the solution admits a supersymmetric limit. 
The present system preserves supersymmetry if there exists a Killing spinor $ \epsilon $ satisfying the following
set of equations~\cite{Boucher:1984yx,Townsend:1984iu,Nozawa:2013maa,Nozawa:2014zia}
\begin{align}
\label{}
\hat \nabla_\mu \epsilon \equiv&\, \big(\nabla_\mu + W(\phi)\gamma_\mu \big)\epsilon=0 \,, \\
\Pi \epsilon \equiv&\, \big(\gamma^\mu \nabla_\mu\phi-8W'(\phi) \big)\epsilon =0 \,. 
\end{align}
Here, $\gamma_\mu$ are gamma matrices satisfying $\{\gamma_\mu, \gamma_\nu\}=2g_{\mu\nu}$. 
The integrability conditions for these equations read
\begin{align}
\label{KSint1}
{\rm det}\left[\hat \nabla_\mu, \hat \nabla_\nu\right]=&\,0 \,, \\
\label{KSint2}
 {\rm det}\Pi =&\,0 \,.
\end{align}
By a straightforward calculation, we find
\begin{align}
\label{}
 {\rm det}\Pi=\frac{16m_0^4(1-\gamma^2)^2f(r)^2}{r^8 f_0(r)^4}\left[
 1-2g^2\sigma \gamma (m_0-\sigma m)(r-m_0-2\sigma m_0 \gamma)
 \right]^2\,. 
\end{align}
Thus, the integrability condition ${\rm det}\Pi=0$ is never satisfied identically for $\gamma^2\ne 1$.\footnote{
By investigating (\ref{KSint1}), one can also verify that the Lorentzian Taub-NUT-AdS breaks supersymmetry.
In the self-dual Euclidean Taub-NUT-AdS case, half of the supersymmetry is preserved~\cite{Nozawa:2017yfl}.
} 
This means that the supersymmetry is completely broken and the hairy 
solution represents the non-Bogomol'nyi-Prasad-Sommerfield (BPS) state. 
This conclusion is consistent with the absence of a degenerate horizon discussed in section~\ref{sec:horizon}, since a BPS black hole must possess an extremal, and hence degenerate, event horizon~\cite{Maeda:2011sh}.

\subsection{Dirac-Misner string}
\label{sec:DMstring}

The present metric exhibits Dirac-Misner string singularities along the
rotation axis $\theta=0,\pi$~\cite{Misner:1963fr,Griffiths:2009dfa}. These singularities can be removed locally by
introducing the coordinates $t\to t\mp2n\gamma\varphi$ in patches around the
north and south poles. Requiring the transition function between the two
patches to be single-valued imposes the periodic identification
\begin{align}
\label{period}
\Delta t=\frac{8\pi n |\gamma|}{k}\,, \qquad k \in \mathbb N\,. 
\end{align}
For $k=1$, one can introduce the coordinate $\psi=t/(2n|\gamma|)$ with period $4\pi$. A constant-$r$ surface therefore corresponds to a ${\rm U}(1)$ bundle over $S^2$ with first Chern number one and has the topology of three-sphere $S^3$. For $k>1$, the corresponding surface has the lens space topology $S^3/\mathbb{Z}_k$ \cite{Jante:2013kha,Boulton:2021wln}.

This bundle structure also provides a natural geometric definition of the
NUT charge. Writing the rotation one-form as $A=2n\gamma \cos\theta \D \phi$ 
and denoting its curvature by $F=\D A$, 
one can define the geometric NUT charge as 
\begin{align}
\label{NUTg}
\mas N \equiv -\frac 1{8\pi} \int_{S^2} F   = n\gamma \,. 
\end{align}

The removal of the Dirac-Misner strings comes, however, at the price of
periodically identifying the time coordinate. Since the stationary Killing
vector $\partial/\partial t$ is timelike in the region
$\Delta_\sigma(r)>0$, its periodically identified orbits give rise to closed timelike
curves. This is the same causal pathology encountered in the ordinary
Taub-NUT-AdS spacetime.

Alternatively, one can advocate the standpoint that the Dirac-Misner string is a physical object carrying 
a distributional singularity, and leave the time coordinate noncompact~\cite{Bonnor,Manko:2005nm}. This viewpoint does not, however,
eradicate the vestiges of closed timelike curves. The norm of the angular Killing vector $\partial/\partial \varphi$ behaves
near the axis ($\theta=0, \pi$) as
\begin{align}
\label{}
g_{\varphi\varphi}=\frac{r^2f_0(r)}{f(r)}\sin^2\theta -4n^2\gamma^2 f(r)\Delta_\sigma (r)\cos^2\theta 
\simeq -4n^2\gamma^2 f(r)\Delta_\sigma (r) \,. 
\end{align}
Since $f(r)>0$ and $\Delta_\sigma(r)>0$ in the asymptotic region $r\to \infty$, the azimuthal orbits become timelike sufficiently close to the axis. It follows that closed timelike curves are unavoidable in the exterior region, even when the
Dirac-Misner strings are retained.

In the geodesic analysis below, we shall return to the periodically identified spacetime when discussing the global extension across the horizon. For the definition of asymptotic charges and Lorentzian thermodynamics, by contrast, we shall retain the Misner strings and keep the time coordinate noncompact.

\subsection{Geodesics}

In this section, we examine the geodesic structure in the present spacetime. 
For related topics in the Taub-NUT-AdS metric, see \cite{Kagramanova:2010bk,Clement:2015cxa}.

\subsubsection{Separability and photon surfaces}

We investigate the geodesic motion governed by 
$p^\mu\nabla_\mu p^\nu=0$ and $p^\mu p_\mu=-\mu^2$, where $p^\mu=\D x^\mu/\D \lambda$, 
$\lambda$ is an affine parameter and $\mu $ is a particle mass. 
To integrate the geodesic equations, it is convenient to employ the
Hamilton-Jacobi formalism. The Hamilton-Jacobi equation takes the form
\begin{align}
\label{}
\frac{\partial \ma S}{\partial \lambda}=\frac 12 g^{\mu\nu}\frac{\partial \ma S}{\partial x^\mu }\frac{\partial \ma S}{\partial x^\nu }\,. 
\end{align}
Consider  the separable form 
\begin{align}
\label{}
\ma S= -\frac 12 \mu^2 \lambda - E t +L \varphi +\ma S_r(r)+\ma S_\theta (\theta ) \,, 
\end{align}
where $E=-p_t$ and $L=p_\varphi $ denote the energy and the angular momentum of the geodesics  respectively.
Inserting this ansatz into the Hamilton-Jacobi equation, we obtain the separable equations
\begin{align}
\label{}
\left(\frac{\D \ma S_\theta }{\D \theta }\right)^2=&\, C-\frac{(L+2n \gamma \cos\theta E)^2}{\sin^2\theta }\,, \\
\left(\frac{\D \ma S_r }{\D r}\right)^2=&\, -\frac{C}{r^2 f_0(r)\Delta _\sigma(r)}+\frac{E^2-\mu^2 f(r)\Delta_\sigma (r)}{f(r)^2\Delta_\sigma(r)^2}\,.
\end{align}
Here, $C(\ge 0) $ represents the separation constant related to the Casimir of ${\rm SU}(2)$ as $C+4n^2\gamma^2 E^2$. 
Equivalently, the Killing tensor $K_{\mu\nu}$ satisfying $K^{\mu\nu}p_\mu p_\nu =C$  and $\nabla_{(\mu}K_{\nu\rho)}=0$ 
is reducible and expressed in terms of symmetrized products of Killing vectors. Note also that the null geodesics with 
$C=0$ correspond to the principal null vectors of the Weyl tensor (\ref{principalnull}). 

The radial equation is therefore reduced to 
\begin{align}
\label{radialeq}
\dot r^2=E^2 -V_{\rm eff}(r) \,, 
\end{align}
where the dot denotes the differentiation with respect to the
affine parameter $\lambda$, and the effective potential $V_{\rm eff}(r)$ is given by 
\begin{align}
\label{Veff}
V_{\rm eff}(r)=\mu^2 f(r)\Delta_\sigma (r)+\frac{Cf(r)^2\Delta_\sigma(r)}{r^2 f_0(r)}\,.
\end{align}

Let us focus here on the null geodesics $\mu=0$. 
First, we consider the ALF case $g=0$ corresponding to the naked singularity, for which the extremum of $V_{\rm eff}(r)$ is realized 
when
\begin{align}
\label{}
(m_0+m)[r-m_0(1+2\gamma)]+(m_0-m)f_0(r)^{2\gamma}[r-m_0(1-2\gamma)]=0\,. 
\end{align}
For $|\gamma|\le 1/2$, we have $r-m_0(1\pm 2\gamma)>0$ in the domain $r>2m_0$. 
Therefore, the extremum of $V_{\rm eff}(r)$ does not exist. 
For $|\gamma|>1/2$ and $n\ne 0$, there always exists a unique maximum of $V_{\rm eff}(r)$ in the domain $r>2m_0$, corresponding 
to the unstable photon orbit. This family of null geodesics constitutes the photon surface~\cite{Virbhadra:1999nm,Claudel:2000yi}.

Let us next consider the photon surface for $g\ne 0$.  For simplicity, 
we concentrate on the case in which the horizon exists (\ref{condhorizon}). 
At the horizon, we have $V_{\rm eff}(r_{\rm h})=0$ and 
\begin{align}
\label{}
 V'_{\rm eff}(r_{\rm h})=\frac{Cf(r_{\rm h})^2}{r_{\rm h}^2 f_0(r_{\rm h})}\Delta_\sigma'(r_{\rm h})>0\,. 
\end{align}
Expanding near infinity $r\to \infty$, we have  $V_{\rm eff}(r)\simeq C [g^2+(1+4g^2n^2\gamma^2)/r^2]>0$, 
and hence
\begin{align}
\label{}
V_{\rm eff}'(r)\simeq -\frac{2C(1+4g^2n^2\gamma^2)}{r^3}<0\,.
\end{align}
Thus, the intermediate value theorem guarantees that there exists at least one local maximum, 
giving rise to the photon surface~\cite{Virbhadra:1999nm,Claudel:2000yi}.

\subsubsection{Null geodesics across the Killing horizon}

Let us finally discuss the behavior of null geodesics near the
nondegenerate horizon at $r=r_{\rm h}$.
For $E>0$, in an angular patch away from the coordinate
singularities on the axis, the geodesic equations give
\begin{align}
\label{nullnearhorizon}
\dot r
=\pm E+O(r-r_{\rm h})\,,
\qquad
\dot t
=\frac{E}{2\kappa(r-r_{\rm h})}+O(1)\,,
\end{align}
where $\kappa=f(r_{\rm h})\Delta_\sigma'(r_{\rm h})/2$ denotes the surface gravity of the horizon, 
see (\ref{kappa}) below.  It follows that
\begin{align}
\label{timehorizon}
t=t_0\pm\frac{1}{2\kappa}
\ln\left|\frac{r-r_{\rm h}}{r_{\rm h}}\right|
+O(r-r_{\rm h})\,,
\end{align}
where $t_0$ is an integration constant 
with the upper and lower signs corresponding to outgoing
and ingoing geodesics, respectively.
This divergence alone does not establish geodesic incompleteness, since $t$ is not a regular coordinate
at the horizon, as is also the case in the Schwarzschild geometry. 

To examine the endpoint of a geodesic in a specified
extension, introduce the ingoing coordinate $v$ by 
$\D v=\D t+\D r/[f(r)\Delta_\sigma(r)]$, in terms of which 
the metric takes the Gaussian-null form 
\begin{align}
\label{ingoingmetric}
\D s^2={}&
-f(r)\Delta_\sigma(r)
\left(\D v+2n\gamma\cos\theta\,\D\phi\right)^2
\notag\\
&+2\left(\D v+2n\gamma\cos\theta\,\D\phi\right)\D r
+\frac{r^2f_0(r)}{f(r)}
\left(\D\theta^2+\sin^2\theta\,\D\phi^2\right),
\end{align}
which is regular at $r=r_{\rm h}$ in a regular angular patch.
Along the geodesics (\ref{timehorizon}), the ingoing null coordinate  behaves 
\begin{align}
\label{vnearhorizon}
v=
\begin{cases}
v_0+O(r-r_{\rm h})\,,
& \text{ingoing}\,,\\[2mm]
v_0+\displaystyle\frac{1}{\kappa}
\ln\left|\frac{r-r_{\rm h}}{r_{\rm h}}\right|
+O(r-r_{\rm h})\,,
& \text{outgoing}\,.
\end{cases}
\end{align}

When the time coordinate is periodically identified,
$v$ inherits the same period $\Delta v=\Delta t=8\pi n|\gamma|/k$ as  (\ref{period}).
Ingoing principal null geodesics have constant $v$ and extend smoothly across the horizon.
Outgoing principal null geodesics, on the other hand, have constant angular coordinates and satisfy
$r-r_{\rm h}=E(\lambda-\lambda_{\rm h})$. Equation (\ref{vnearhorizon}) therefore shows that they
wind infinitely many times around the periodic $v$ direction as $\lambda\to\lambda_{\rm h}$.
They accumulate on a closed null generator of the horizon rather than approaching a single point.
Thus, these geodesics have no continuous extension through the horizon in the chosen ingoing extension, which is
consequently null geodesically incomplete. Choosing instead the outgoing extension interchanges the roles of the two families, giving rise to two inequivalent extensions across the horizon~\cite{Misner:1963fr,Hawking:1973uf}.
There is no canonical prescription that singles out either extension as preferred. Alternatively, both families of null geodesics can be continued across the horizon by adjoining the ingoing and outgoing extensions simultaneously,
but the resulting spacetime is non-Hausdorff~\cite{Miller:1971em}.

\subsection{Asymptotic structure and conserved charges}

In this section, we examine the asymptotic structure of spacetime and demonstrate that the
solution is asymptotically locally AdS. Here, we allow the Dirac-Misner string and 
do not impose the periodic identification of the time coordinate. 

For this purpose, it is advantageous to work with 
the areal radius of the orbit space of the stationary Killing vector
\begin{align}
\label{}
S\equiv g_{\theta\theta}^{1/2}=r\sqrt{\frac{f_0(r)}{f(r)}}\,, 
\end{align}
in terms of which 
the scalar field is expanded asymptotically ($S\to \infty$) as 
\begin{align}
\label{asyexscalar}
\phi \simeq \frac{\phi_1}{S}+\frac{\phi_2}{S^2} \,, 
\end{align}
where 
\begin{align}
\label{}
\phi_1=-\frac{4\sigma \alpha }{1+\alpha^2}m_0 \,, \qquad 
\phi_2=m \gamma \phi_1 \,. 
\end{align}
This asymptotic fall-off behavior of the scalar field (\ref{asyexscalar}) is determined by the mass term (\ref{masseig}) of the scalar field. 
The two modes of the scalar field are not independent but functionally related  by 
\begin{align}
\label{phi2eq}
\phi_2=\gamma \phi_1 \sqrt{\frac{\phi_1^2}{4(1-\gamma^2)}-n^2} \,.
\end{align}
Writing $\phi_2 =\D \ma W/\D \phi_1$, we find
\begin{align}
\label{}
\ma W(\phi_1)=\frac{4\gamma(1-\gamma^2)}{3}\left(\frac{\phi_1^2}{4(1-\gamma^2)}-n^2\right)^{3/2}\,,
\end{align}
This function $\ma W$ defines the boundary condition in designer gravity, corresponding to a multi-trace deformation of the dual conformal field theory \cite{Hertog:2004ns,Anabalon:2015xvl}.

Similarly, the metric can be expanded asymptotically as
\begin{align}
\label{}
\D s^2 \simeq& \, -\left(1+4g^2n^2\gamma^2 -\frac{2M_1}{S}+g^2S^2\right) \left(\D t+2n\gamma \cos\theta \D \varphi\right)^2 
\notag \\
&\, +\frac{\D S^2}{1+\Gamma +g^2S^2-2M_2/S} +S^2 \left(\D \theta^2+\sin^2\theta \D \varphi ^2\right)\,, 
\label{asyexmetric}
\end{align}
where 
\begin{align}
\label{Gamma}
\Gamma =&\,3g^2n^2 \gamma^2+\frac 14 g^2 \phi_1^2 \,, \\ 
M_1=&\, m\gamma -\frac 23 g^2 \gamma\left[6m^3 \gamma^2+m m_0^2 (1-10\gamma^2)
+\sigma m_0^3(4\gamma^2-1) \right]\,,\\
M_2=&\, M_1-\frac 13 g^2 \phi_1 \phi_2 \,.
\end{align}
The unfamiliar terms proportional to $g^2n^2$ in $g_{tt}$ and $g_{SS}^{-1}$ 
also appear from the asymptotic expansion of the Taub-NUT-AdS metric (\ref{TNAdS}). 
The slowly decaying scalar field backreacts on the metric through the extra contribution to $\Gamma$ and $M_2$.

We now identify the mass of the spacetime with the  Hamiltonian charge associated with 
time translation  $\xi=\partial/\partial t$. According to the covariant phase space method~\cite{Iyer:1994ys}, 
the variation of the Hamiltonian $H_\xi$ is given by 
\begin{align}
\label{deltaH}
\delta H_\xi=\int_{S^\infty} \delta \boldsymbol Q-i_\xi \boldsymbol \theta\,,
\end{align}
where $i_\xi $ is an interior product contracting the first index with $\xi^\mu$, and 
$\boldsymbol Q$ represents the Noether charge 2-form 
\begin{align}
\label{}
\boldsymbol Q_{\mu\nu}= &\,-\frac 1{16\pi} \epsilon_{\mu\nu\rho\sigma}\nabla^\rho \xi^\sigma \,,
\end{align}
and 
$\boldsymbol \theta =\boldsymbol \theta^{(\rm grav)}+\boldsymbol \theta^{(\varphi)}$ is the symplectic potential 
\begin{align}
\boldsymbol \theta _{\mu\nu\rho}^{(\rm grav)}=&\,\frac 1{16\pi} \epsilon_{\sigma\mu\nu\rho}g^{\sigma\tau}g^{\kappa\lambda}
(\nabla_\kappa \delta g_{\tau\lambda}-\nabla_\tau \delta g_{\kappa\lambda}) \,,
\\
\boldsymbol \theta _{\mu\nu\rho}^{(\rm \phi)}=&\,-\frac 1{16\pi} \epsilon_{\sigma\mu\nu\rho}\delta\phi \nabla^\sigma \phi \,. 
\end{align}
Here, the integration surface $S^\infty$ is taken to be the two-sphere at infinity in the
orbit space of the stationary Killing vector $\xi=\partial/\partial t$. 
The surface integral is understood in terms of pull-backs by local sections of the time fibration, following the
construction of \cite{Bossard:2008sw}.

To evaluate its variation, one must specify the covariant phase space, namely the class of asymptotic configurations and variations allowed by the boundary conditions.
In asymptotically locally AdS gravity, these conditions fix the conformal class of the boundary metric, together with the boundary condition imposed on the scalar field. 
From the asymptotic form (\ref{asyexmetric}) of the metric, the boundary 
metric reads 
\begin{align}
\label{boundary}
\D \bar s_3^2=-  \left(\D t+2n\gamma \cos\theta \D \varphi\right)^2 +g^{-2}\left(\D \theta^2+\sin^2\theta \D \varphi ^2\right) \,. 
\end{align}
The boundary metric is deformed from the cylinder and not conformally flat. 
Thus, the spacetime does not obey the asymptotically AdS boundary conditions preserving ${\rm SO}(3,2)$ asymptotic symmetry \cite{Hertog:2004dr,Henneaux:2006hk}. 
In the presence of the NUT charge, it is reasonable to consider boundary conditions that preserve the 
boundary data (\ref{boundary}), which we shall assume hereafter. Under these relaxed boundary conditions, 
only the isometry group ${\mathbb R}\times {\rm SU}(2)$ of the boundary metric is preserved.

Regarding $n$ as fixed boundary data and restricting the phase space to variations satisfying
$\delta n=0$,  the physical conserved mass can be computed by (\ref{deltaH}). 
Substituting the asymptotic expansions (\ref{asyexscalar}) and (\ref{asyexmetric}) and  
using the relation (\ref{Gamma}), we find  that the surface integral is indeed finite, giving\footnote{
If we vary $n$, the integrand of the expression $\delta H_\xi $ involves the term proportional to $ r \delta n$, 
which prevents the Hamiltonian charge from converging.}
\begin{align}
\label{}
\delta H_\xi =\delta M_2+\frac 14 g^2 \left(2\phi_2 \delta \phi_1+\phi_1\delta \phi_2 \right)
=\delta M_1+\frac 1{12}g^2\left(2\phi_2\delta \phi_1-\phi_1 \delta \phi_2\right)\,.
\end{align}
This expression agrees with the static case \cite{Liu:2013gja}. 
For fixed $n$,  
this variation is integrable by virtue of (\ref{phi2eq}) and gives the mass $M=H_\xi$ as 
\begin{align}
\label{Mass}
M=M_1+\frac 13 g^2 m n^2 \gamma(\gamma^2-1)\,,
\end{align}
where we have set to zero the undetermined term that may depend on
$g$, $\gamma$, and $n$. 

When the NUT parameter vanishes $n=0$, we obtain the 
well-known result $M=M_1$ \cite{Hertog:2004dr,Henneaux:2006hk}. 
When the scalar contribution vanishes ($|\gamma|=1$), we have 
$M=\ti m$ in (\ref{tirm}), again reproducing the well-established result. 
These two limiting cases provide nontrivial consistency checks of the mass formula obtained above.

\subsection{Black hole thermodynamics}

We now discuss the thermodynamics of the solutions admitting a regular
Killing horizon at $r=r_{\rm h}$, with parameters satisfying (\ref{condhorizon}).
We adopt the Lorentzian area-law prescription of~\cite{Hennigar:2019ive}, 
retaining the Dirac-Misner strings and leaving the time coordinate noncompact.
The horizon area is given by
\begin{align}
\label{area}
A_{\rm h}=4\pi S(r_{\rm h})^2=4\pi r_{\rm h}^2 \frac{f_0(r_{\rm h})}{f(r_{\rm h})} \,.
\end{align}
The surface gravity of the horizon associated with the Killing vector
$\partial/\partial t$ is computed as 
\begin{align}
\label{kappa}
\kappa =\frac 12 f(r_{\rm h})\Delta_\sigma'(r_{\rm h})\,. 
\end{align}
The temperature and the entropy of the black hole are given by
\begin{align}
\label{}
T=\frac{\kappa}{2\pi}\,, \qquad S_{\rm BH}=\frac{A_{\rm h}}{4} \,. 
\end{align}

Here, we wish to derive the first law, involving the mass derived in (\ref{Mass}). 
Since this mass was obtained on a phase space with fixed NUT parameter (otherwise the integrand of $\delta H_\xi$
diverges), 
we restrict the thermodynamic variations considered here to $\delta n=0$.
The theory parameters $g$ and $\gamma$ are also held fixed, and variations 
are taken within a fixed branch $\sigma$.
Thus, the variations remain within the same boundary metric and scalar
mixed boundary condition used in defining the mass.
Any additive function of the fixed parameters in (\ref{Mass})
does not affect the mass variation considered below.

With these remarks in mind, one finds that the horizon contribution alone
does not reproduce the mass variation
$\delta M\neq (\kappa/8\pi)\delta A_{\rm h}$.
This feature is already present in the area-law formulation of
ordinary Taub-NUT-AdS thermodynamics, where the Misner strings
provide an additional contribution \cite{Hennigar:2019ive}.
Following the prescription in \cite{Hennigar:2019ive} and its geometric interpretation
in \cite{Bordo:2019tyh,BallonBordo:2019vrn} (see also 
\cite{Hawking:1998ct,Mann:2004mi,Godazgar:2022jxm,Awad:2022jgn,Liu:2022wku,Wu:2023fcw,Liu:2023uqf} for related topics), we introduce the Misner potential
\begin{align}
\label{}
\psi=\frac 1{8\pi n \gamma }\,, 
\end{align}
which is associated with the surface gravity for 
the Killing horizons at $\theta=0, \pi$ with generator
\begin{align}
\label{}
\xi'=\frac{\partial}{\partial t}\mp\frac{1}{2n\gamma} \frac{\partial}{\partial \varphi}\,. 
\end{align}
Since the two strings at the north and south poles are symmetrically distributed in the present
solution, their contributions can be combined into a single total
Misner charge $\ma N$ with conjugate potential $\psi$.

We define the thermodynamic Misner charge by requiring the
restricted first law
\begin{align}
\label{firstlaw}
\delta M=T \delta S _{\rm BH}
+\psi\,\delta\ma N\,,
\end{align}
with $\delta n=\delta g=\delta\gamma=0$. 
Since $\psi$ has already been fixed, this relation determines the $m$ dependence of $\ma N$, up to an additive function of the fixed parameters.

The resulting Misner charge admits a geometric interpretation in terms of a renormalized Komar-type integral 
over tubes surrounding the Dirac-Misner strings, following \cite{Bordo:2019tyh,BallonBordo:2019vrn}.
We write this schematically as
\begin{align}
\label{Misnertube}
\ma N
=\frac{1}{16\pi\psi}
\left(
\int_{\cal T}\star\D\xi-\Omega^{(r=\infty)}
\right)\,,
\end{align}
where $\cal T$ denotes the appropriately oriented union of the tubes surrounding
the north and south Misner strings on a constant-$t$ hypersurface, extending 
from $r=r_{\rm h}$ to infinity.
The second term $\Omega^{(r=\infty)}$ in (\ref{Misnertube}) denotes a renormalized Komar-type representation
 that removes the asymptotic AdS divergence of the Komar-type tube integral.
In Einstein-AdS gravity, such a subtraction is naturally implemented by a Killing co-potential~\cite{Bordo:2019tyh,BallonBordo:2019vrn}.
Here we fix the corresponding subtraction by requiring finiteness and 
agreement with the standard Taub-NUT-AdS result in the scalar-free limit.
For the present solution, the tube integral can be expressed in terms of the auxiliary function
\begin{align}
\label{Primitive}
P_\sigma (r) \equiv\, &\frac {\Delta _\sigma (r)}{2\gamma (m_0+m)}\Biggl(
f(r)f_0(r)^\gamma -\frac{m_0(1-\sigma )}{m_0-m}\Biggr)\notag \\
&-\frac{\sigma}{2\gamma(m_0+m \sigma)}\Big\{1+g^2 \left[(r-(1+2\sigma \gamma)m_0)^2+(4\gamma^2-1)m_0^2
-6\sigma m(m_0+\sigma m)\gamma^2 \right]\Big\}
\,,
\end{align}
satisfying 
\begin{align}
\label{Pr}
\frac{\D P_\sigma}{\D r}=\frac{f(r)^2 \Delta_\sigma(r)}{r^2 f_0(r) }\,, \qquad 
P_\sigma (r)-g^2 S(r)=O(1/r)\,, 
\end{align}
where $S(r)=r\sqrt{f_0(r)/f(r)}$ is the areal radius of the orbit space.
The second relation specifies the additive constant of the primitive (\ref{Primitive}) such that 
its finite part vanishes at infinity when the divergent term is expressed in terms of this areal radius.
Accordingly, the renormalized tube integral takes the form 
\begin{align}
\label{Misnerchargeintegral}
\ma N
=-8\pi(n\gamma)^3
\lim_{R\to\infty}
\left[
\int_{r_{\rm h}}^R
\frac{f(r)^2\Delta_\sigma(r)}{r^2f_0(r)}\D r
-g^2S(R)
\right]\,.
\end{align}
Using (\ref{Pr}), this immediately gives 
\begin{align}
\label{Misnercharge}
\ma N=8\pi(n\gamma)^3P_\sigma(r_{\rm h})\,.
\end{align}
One can then verify directly that (\ref{Misnercharge}) indeed satisfies the restricted first law (\ref{firstlaw}).
When the scalar field vanishes $\gamma^2=1$, this expression reduces to 
$\ma N=-4 \pi n^3[1+3g^2(n^2-\ti r_{\rm h}^2)]/\ti r_{\rm h}$, where 
$\ti r_{\rm h}$ is the horizon locus for the radial variable in (\ref{tirm}). 
This recovers the expression for the Taub-NUT-AdS spacetime~\cite{Bordo:2019tyh,BallonBordo:2019vrn}.

It is important to distinguish the thermodynamic Misner charge
$\ma N$ from the geometric NUT charge $\mas N=n\gamma$ in (\ref{NUTg}).
In particular, fixing $n$ does not imply $\delta\ma N=0$, since
$\ma N$ depends nontrivially on $m$ and the horizon data.
Consequently, the Misner contribution in (\ref{firstlaw}) remains
necessary even for variations at fixed NUT parameter.

The construction above establishes a thermodynamic prescription
on the restricted phase space with fixed $n$.
Allowing $n$ to vary also changes the scalar mixed boundary
condition and requires a consistent extension of the charge
definitions and their normalization beyond this phase space.
We do not address that extension here. A complete thermodynamic
formulation allowing independent variations of $m$ and $n$
(and $g$ and $\gamma$) 
is left for future work.

\section{Euclidean solution}
\label{sec:Euc}

Euclidean gravitational instantons are of primary importance due to
their relevance for non-perturbative effects in quantum gravity \cite{Hawking:1976jb,Gibbons:1979xm}. 
Euclidean supersymmetric solutions have attracted particular interest in the context of localization techniques~\cite{BenettiGenolini:2019jdz,BenettiGenolini:2024xeo}. Motivated by these developments, this section examines
 the Euclidean continuations of the Lorentzian solutions discussed in the main text.

\subsection{Wick-rotated solution}

By analytic continuation $t\to i t$, $n\to i n$ of (\ref{metric}), 
one can obtain a solution with Euclidean signature 
\begin{align}
\label{metricE}
\D s^2=&\,f(r) \Delta_\sigma (r) \left(\D t+2 n \gamma \cos\theta \D \varphi \right)^2
+\frac 1{f(r)}\left[\frac{\D r^2}{\Delta_\sigma (r)}+r^2 f_0(r) \left(\D \theta ^2+\sin^2\theta \D \varphi^2\right)\right]\,, 
\\
\label{scalarE}
\phi =&\, \sigma \frac{2\alpha}{1+\alpha^2}\ln f_0(r) \,, 
\end{align}
where $f_0$, $f$ and $\Delta_\sigma$ are given by (\ref{f0}), (\ref{f}) and (\ref{Delta}), while 
the parameter $m_0$ is now understood to be 
\begin{align}
\label{constraintEuc}
m_0=\sqrt{m^2-n^2} \,. 
\end{align}
This solution obeys the Euclidean counterpart of the field equations derived from the Lagrangian (\ref{Lag}). 

One sees that (\ref{constraintEuc}) requires $|m|>|n|$. Under the constraint (\ref{constraintEuc}), 
the scalar curvature  takes the same form as in the Lorentzian case (\ref{Ricciscalar}). 
Since $f$ is not positive-definite, the singularity may occur where $f(r)^{-1}$ vanishes,
in addition to $r=0$ and $r=2m_0$.

\subsubsection{Ambi-K\"ahler structure}

The Euclidean continuation possesses a richer geometric structure than is apparent from its form. To make this structure explicit, 
let us define two-forms 
\begin{align}
\label{}
\omega_\pm =(\D t+2n \gamma \cos\theta \D \varphi ) \we \D r \pm r^2\frac{f_0(r)}{f(r)}\sin\theta \D \theta \we \D \varphi \,,
\end{align}
where the $\pm$ sign has nothing to do with the branch but corresponds to the orientation $\star \omega_\pm =\pm \omega_\pm$. 
Writing the almost complex structures as $(J_\pm)_\mu{}^\nu=(\omega_\pm)_{\rho\mu} g^{\rho \nu}$ satisfying 
$(J_\pm)_\mu{}^\rho(J_\pm)_\rho{}^\nu = -\delta_\mu{}^\nu$, 
one can check that the Nijenhuis tensors for $J_\pm $ vanish. 
It follows that these complex structures are integrable~\cite{Klemm:2013eca,Nozawa:2015qea}. 
Nevertheless, the tensors $\omega_\pm$ are not covariantly constant, 
which implies that they do not define the K\"ahler structures.

By a direct calculation, one finds that the exterior derivative of $\omega_\pm$ obeys
\begin{align}
\label{}
\D \omega_\pm -\theta_\pm \we \omega_\pm=0 \,.
\end{align}
Here, $\theta_\pm$ are Lee forms~\cite{Gauntlett:2003cy}, which turn out to be exact 
\begin{align}
\label{}
\theta _\pm =-2 \D \left( \ln \Omega_\pm \right)\,, 
\end{align}
where 
\begin{align}
\label{}
\Omega_\pm =\left(\frac{2m_0f_0(r)^{\gamma-1}}{|m+m_0|r^2 (1\mp Z)^2}\right)^{1/2}\,, \qquad 
Z\equiv \frac{n f_0(r)^\gamma}{m+m_0}  \,. 
\end{align}
Therefore, the conformal transformations $\hat g^\pm _{\mu\nu}=\Omega_\pm^2 g_{\mu\nu}$ allow us to 
obtain the K\"ahler manifolds ($M_\pm$, $\hat g^\pm$, $\hat \omega_\pm$) with K\"ahler forms $\hat \omega_\pm \equiv \Omega_\pm^2 \omega_\pm $~\cite{Gauntlett:2003cy}. Specifically, 
the conformal class of the metric (\ref{metricE}) admits an ambi-K\"ahler structure,
defining conformally related but oppositely oriented 4-dimensional K\"ahler metrics
\cite{Apostolov:2013oza,Keaton}. 
This structure places the Euclidean family within a geometrically distinguished class closely related to familiar four-dimensional gravitational instantons.

\subsubsection{Bolts and nuts}

We first investigate whether the geometry admits a regular bolt at $r=r_{\rm b}$, 
where 
\begin{align}
\label{bolt}
\Delta _\sigma (r_{\rm b})=0 \,, \qquad 0<f(r_{\rm b})<\infty \,, \qquad f_0(r_{\rm b})>0 \,. 
\end{align}
Obviously, we need $r_{\rm b}>2m_0 >0$. 
The absence of a conical singularity then requires the Euclidean time
coordinate to have the period 
\begin{align}
\label{periodtE}
t\sim t+\beta \,, \qquad \beta =\frac{2\pi}{\kappa_{\rm E}} \,,
\end{align}
where $\kappa_{\rm E}=f(r_{\rm b}) \Delta_\sigma '(r_{\rm b})/2$, see (\ref{kappa}). 
This period must be compatible with the period $\Delta t=8\pi |n\gamma|/k$ ($k\in \mathbb N$), which is required to remove
the Dirac-Misner string along the axis, see (\ref{period}). 
This yields 
\begin{align}
\label{regularitybolt}
k=4g^2 |n\gamma| f_0(r_{\rm b})^{-\sigma \gamma}[r_{\rm b}-(2\sigma\gamma+1)m_0] \,.
\end{align}

For definiteness of the argument, we confine ourselves to the $\gamma=\pm 1/2$ case in what follows, 
for which the ${\cal N}=8$ embedding is possible. In these cases, two $\sigma$-branches coincide and 
\begin{align}
\label{}
f(r)=\frac{\sqrt{r(r-2m_0)}}{w(r)}\,, \qquad 
\Delta_\sigma(r)=1+g^2 (w(r)^2-n^2)\,, \qquad 
w (r)\equiv r-m_0+{\rm sgn}(\gamma ) m \,. 
\end{align}
When ${\rm sgn}(\gamma)m>0$, we have 
$w (r)>m_0+{\rm sgn}(\gamma)m=m_0+|m|>|n|$ throughout $r>2m_0$, and hence 
$\Delta _\sigma (r)>0$. Thus, no bolt exists for this sign choice. 
Therefore, the existence of a bolt requires
\begin{align}
\label{boltex1}
g^2n^2>1 \,, \qquad {\rm sgn}(\gamma)m<0 \,. 
\end{align}
Under these conditions,  the locus of bolt is 
\begin{align}
\label{}
r_{\rm b}=m_0-{\rm sgn}(\gamma ) m+w_{\rm b}\,, \qquad 
 w_{\rm b}=\sqrt{n^2-g^{-2}}>0 \,.
\end{align}
For ${\rm sgn}(\gamma)m<0$, the zero of $w(r)$ occurs at
\begin{align}
\label{}
r_{\rm s}=m_0-{\rm sgn}(\gamma)m=m_0+|m|>2m_0\,,
\end{align}
where $f(r)$ diverges and the curvature becomes singular.
Due to $r_{\rm b}=r_{\rm s}+w_{\rm b}$, the bolt is always located outside this singular surface, 
and hence $r_{\rm b}>r_{\rm s}>2m_0$. 
It follows that  the regularity condition for the bolt (\ref{regularitybolt}) reduces to 
\begin{align}
\label{}
k=2g^2|n|\sqrt{r_{\rm b}(r_{\rm b}-2m_0)}>2 (gn)^2>2 \,,
\end{align}
where we have used $r_{\rm b}(r_{\rm b}-2m_0)=(|m|+w_{\rm b})^2-m_0^2=n^2+2|m|w_{\rm b}+w_{\rm b}^2>n^2$. 
This equation means that a regular bolt requires the constant-$r$ hypersurfaces
to have lens-space topology $S^3/\mathbb Z_k$ with $k\geq 3$.\footnote{
The lens space $S^3/\mathbb Z_k$ itself admits a spin structure for every $k\in\mathbb N$. 
By contrast, the four-dimensional Euclidean bolt fillings, whose topology is that of the total space of
$\mathcal O(-k)\to S^2$, admit a spin structure if and only if $k$ is even.
This distinction is relevant to their global interpretation as 
supergravity backgrounds. See appendix D.1 of \cite{Martelli:2012sz} for details.
}
To verify that regular bolt solutions actually exist for every $k\geq3$,
we introduce the dimensionless variables $\mathsf q\equiv g^2n^2>1$ and $\mathsf m\equiv |m|/|n|>1$.
The regularity condition for the bolt (\ref{regularitybolt}) is rewritten as 
\begin{align}
\label{}
\mathsf m=\frac{k^2+4(1-2\mathsf q)\mathsf q}{8\mathsf q^{3/2}\sqrt{\mathsf q-1}} \,. 
\end{align}
Combined with $k>2$, the condition $\mathsf m>1$ amounts to 
\begin{align}
\label{}
1<\mathsf q<\frac{k^2}{4(k-1)} \,. 
\end{align}
For every integer $k\ge 3$, one can always find $\mathsf q$ in this interval. 
Consequently, there exists a one-parameter family of regular bolt solutions for each $k\geq3$,
parametrized by $q$ in the above interval, with the sign of $m$ chosen so that
${\rm sgn}(\gamma)m<0$.

Let us next consider a nut for $|\gamma|=1/2$, corresponding to an isolated fixed point of the Killing vector $\partial/\partial t$.
This is realized when the zero of $f(r)^{-1}$ coincides with the double zero of $\Delta_\sigma(r)$,
i.e., we need ${\rm sgn}(\gamma)m<0$ and 
\begin{align}
\label{}
g^2n^2=1 \,, \qquad  r_{\rm n}=m_0-{\rm sgn}(\gamma) m \,.
\end{align}
In this case, we obtain
\begin{align}
\label{}
\Delta_\sigma(r)=g^2(r-r_{\rm n})^2 \,, \qquad \frac{1}{f(r)}=\frac{r-r_{\rm n}}{\sqrt{r(r-2m_0)}}\,, 
\end{align}
satisfying $r_{\rm n}>2m_0$ and $r_{\rm n}(r_{\rm n}-2m_0)=n^2$. 
Introducing
\begin{align}
\label{}
\psi =\frac{t}{2n\gamma}\,, 
\end{align}
the periodic identification required to remove the Dirac-Misner strings implies
\begin{align}
\label{}
\Delta\psi=\frac{\Delta t}{|n|}=\frac{4\pi}{k} \,.
\end{align}
In terms of $\psi$, the metric is boiled down to 
\begin{align}
\label{metricSO4}
\D s^2=\frac{\D r^2}{g^2 (r-r_{\rm n})\sqrt{r(r-2m_0)}}+(r-r_{\rm n})\sqrt{r(r-2m_0)}\left[(\D \psi+\cos\theta \D \varphi)^2 +
\D \theta^2+\sin^2 \theta \D \varphi^2 
\right]\,. 
\end{align}
In the vicinity of the nut $r\simeq r_{\rm n}$, 
we find that the scalar field is finite at the nut and the metric is approximated by 
\begin{align}
\label{nutmetric}
\D s^2\simeq \D \rho^2+\frac{\rho^2}{4}\left[(\D \psi+\cos\theta \D \varphi)^2+\D \theta^2+\sin^2\theta \D \varphi^2\right]\,,
\end{align}
where we have defined $\rho^2=4|n|(r-r_{\rm n})$. 
For $k=1$, the angular metric is the standard round metric on $S^3$,
and $\rho=0$ is a smooth nut with a neighborhood locally modeled on $\mathbb R^4$. 
For $k>1$, the nut instead has an orbifold singularity locally modeled
on $\mathbb R^4/\mathbb Z_k$~\cite{Cvetic:1998hg,Martelli:2012sz}.
Thus, a smooth nut requires $k=1$ and the global geometry admits an enhanced symmetry ${\rm SO}(4)$. 
The metric (\ref{metricSO4}) corresponds to the Einstein-frame metric of the spherically symmetric conformal scalar instanton studied
in \cite{deHaro:2006ymc}. Our construction thus recovers this known instanton as a regular
nut limit of the NUT-charged family.

\subsection{An exceptional Euclidean solution for $\gamma=0$ ($\alpha=1$)}

Since  the $\gamma=0$ solution is reduced to a static configuration in the Lorentzian case, 
the NUT parameter drops out. The same is true for the above Euclidean metric (\ref{metricE}).

Nevertheless, 
we can construct, in the Euclidean case, yet another solution for $\gamma=0$:
\begin{align}
\label{metricE2}
\D s^2=&\,\check f(r) \check \Delta (r) \left(\D t+2 n_0  \cos\theta \D \varphi \right)^2
+\frac 1{\check f(r)}\left[\frac{\D r^2}{\check \Delta (r)}+r^2 f_0(r) \left(\D \theta ^2+\sin^2\theta \D \varphi^2\right)\right]\,, 
\\
\label{scalarE2}
\phi =&\, \pm \ln f_0(r) \,, 
\end{align}
where $f_0(r)=1-2m_0/r$ as before, and 
\begin{align}
\label{}
\check f(r)=&\, \frac{m_0}{m_0-n_0 \ln f_0(r)} \,, \\
\check \Delta(r)=&\, 1+g^2 \left[r^2-2(m_0-n_0)r-2m_0 n_0 -\frac{n_0}{m_0}r^2 f_0(r) \ln f_0(r)\right]\,.
\end{align}
Here, $m_0$ and $n_0$ are unconstrained free parameters. 
The plus/minus sign in the scalar field (\ref{scalarE2}) does not label distinct branches, but merely reflects a sign convention, since the potential 
with $\gamma=0$ is an even function of $\phi$ as 
\begin{align}
\label{}
V(\phi)=-g^2(2+\cosh\phi)\,.
\end{align}
Although the combination $\D t+2n_0\cos\theta\D\varphi$ can be Wick rotated by analytically continuing $t $ and $n_0$ simultaneously, the metric functions depend linearly on $n_0$ rather than only through $n_0^2$. 
This obstructs a real Lorentzian continuation.

It is worth commenting that 
the solution (\ref{metricE2}) is obtainable from (\ref{metricE}) by taking the scaling limit
\begin{align}
\label{}
\gamma \to 0 \,, \qquad m\to \infty\,, \qquad n\to \infty\,, 
\end{align}
while keeping $m_0$ and $n_0\equiv n\gamma$ fixed. Here, $m=\sqrt{m_0^2+n^2}\sim n$ in this limit. This may be viewed as an infinite-boost limit along the hyperbola $m^2-n^2=m_0^2$. Such a limit has no Lorentzian analogue, where the corresponding parameters instead obey a relation involving $m^2+n^2$.

The solution is ALF for $g=0$ and asymptotically $\mathbb H^4$ for $g\ne 0$, 
and possesses the conformally K\"ahler structure as (\ref{metricE}). 
The solution admits a discrete symmetry
\begin{align}
\label{}
r\to 2m_0-r\,,\qquad
n_0\to -n_0\,,\qquad
\varphi\to-\varphi\,,\qquad
\phi\to-\phi\,,
\end{align}
where we have used $f_0(2m_0-r)=f_0(r)^{-1}$.
This transformation exchanges the two disconnected regions $r>2m_0$ and $r<0$.
Hence, once one of these asymptotic regions is selected, the two signs 
of $n_0$ are no longer equivalent within that region. 

The scalar curvature is written as
\begin{align}
\label{}
R=\frac{2m_0^2\check f(r)\check \Delta(r)}{r^4f_0(r)^2}-\frac{4g^2}{f_0(r)}\left(3-\frac{6m_0}{r}+\frac{2m_0^2}{r^2}\right)\,, 
\end{align}
which is formally obtained by setting $\gamma=0$ in (\ref{Ricciscalar}). Thus, the surface 
$\check \Delta(r)=0$ is regular, and the singularity may occur where $\check f(r)^{-1}=0$, 
besides $r=0$ and $r=2m_0$. 

We now consider the case $m_0>0$ and focus on the region outside the singular surface $r=2m_0$. 
Unlike the generic Euclidean continuation (\ref{metricE}), the exceptional solution admits a particularly simple characterization of the bolt structure, which we analyze below.
To examine whether a bolt, defined by $\check \Delta(r_{\rm b})=0$, can occur in this region, it is useful to note the relation 
\begin{align} 
\label{} 
\check f(r)\check \Delta'(r)=2g^2(r-m_0)\,. 
\end{align} 
In the region $r>2m_0$, one has $\ln f_0(r)<0$ hence $\check f(r)>0$ for $n_0\geq 0$.
Thus, $\check\Delta'(r)>0$ holds throughout $r>2m_0$. Together with $\check \Delta(2m_0)=1+2g^2m_0 n_0>0$, no bolt exists for $n_0\geq0$.

Let us next consider the case $n_0<0$. Then, there exists a surface $r=r_{\rm s}$ 
at which $\check f^{-1}$ vanishes
\begin{align} 
\label{rsEuc} 
\ln f_0(r_{\rm s})=\frac{m_0}{n_0}\quad 
\Longleftrightarrow \quad 
r_{\rm s}=m_0\left[1+\coth\left( \frac{m_0}{2|n_0|}\right)\right]\,.
\end{align}
In view of  $\coth x>1$ for $x>0$, we obtain $r_{\rm s}>2m_0$.
Since $\check f^{-1}$ changes sign from negative to positive across $r=r_{\rm s}$, 
the above relation implies that $\check\Delta(r)$ decreases for $2m_0<r<r_{\rm s}$ and increases for $r>r_{\rm s}$. Thus, $r=r_{\rm s}$ is the unique minimum of $\check\Delta(r)$ in the region $r>2m_0$. 
At this surface, 
\begin{align} 
\label{}
\check \Delta(r_{\rm s}) =1-2g^2|n_0|(r_{\rm s}-m_0)= 1-2g^2m_0 |n_0| \coth\left(\frac{m_0}{2|n_0|}\right)\,. 
\end{align} 
It follows that an outer bolt $r_{\rm b}>r_{\rm s}>2m_0$ exists if and only if 
\begin{align} 
\label{boltex2} 
2g^2m_0 |n_0| \coth\left(\frac{m_0}{2|n_0|}\right)>1\,, \qquad n_0<0\,. 
\end{align}
In particular, the ALF case ($g\to 0$) admits no bolt. 

For the regularity of the bolt, the time coordinate must have the period (\ref{periodtE}), where 
the surface gravity is now given by $\kappa_{\rm E}=\check f(r_{\rm b})\check \Delta '(r_{\rm b})/2$.  
Equating this period with $\Delta t=8\pi |n_0|/k$ ($k\in \mathbb N$), which is required to remove
the Dirac-Misner string along the axis, 
we obtain the following  condition
\begin{align}
\label{}
r_{\rm b}=m_0+\frac{k}{4g^2|n_0|} \,. 
\end{align}
Together with $r_{\rm b}>r_{\rm s}$ and the existence condition of the bolt (\ref{boltex2}),  
we need 
\begin{align}
\label{}
k>4g^2|n_0|(r_{\rm s}-m_0) >2 \,. 
\end{align}
As in the $|\gamma|=1/2$ cases, 
a lens-space identification $S^3/\mathbb Z_k$ with $k\geq 3$ is 
necessary for the existence of a regular bolt, and $k$ must be even to
admit a spin structure for the total space. 

To show that regular bolt solutions indeed exist for every integer $k\geq 3$, we introduce the dimensionless variable
$x_b\equiv (r_{\rm b}-m_0)/m_0>1$. The bolt equation $\check\Delta(r_{\rm b})=0$ and the periodicity condition above can then be rewritten as
\begin{align}
\label{boltcond}
m_0 |n_0|=\frac{k}{4g^2 x_{\rm b}}>0 \,, \qquad 
\frac{m_0}{|n_0|}=\ln \left(\frac{x_{\rm b}+1}{x_{\rm b}-1}\right)+\frac{2(k-2)x_{\rm b}}{k(x_{\rm b}^2-1)} >0  \,. 
\end{align}
For fixed $g\ne 0$, any choice of $x_{\rm b}>1$ and integer $k\ge 3$ therefore determines
a unique pair of positive values $m_0$ and $|n_0|$.
Moreover, the  inequality $m_0/|n_0|> \ln [(x_{\rm b}+1)/(x_{\rm b}-1)]$ implies 
$x_{\rm b}>\coth[m_0/(2|n_0|)]$, and hence $r_{\rm b}>r_{\rm s}$. Thus, the zero at $r=r_{\rm b}$ lies in the regular region where
$\check f>0$, and represents an outer bolt. Consequently, for each integer $k\geq 3$, there exists a one-parameter family of regular bolt solutions parametrized by $x_{\rm b} (>1)$. This establishes that $k\geq 3$ is not only necessary but also sufficient for the existence of regular bolt solutions, with the continuous parameters constrained by (\ref{boltcond}), as we desired to show.

Finally, we consider the possibility of a nut, for which 
the divergence of $\hat f$ is compensated by the double zero of $\check \Delta$. 
Inserting (\ref{rsEuc}) into $\check \Delta (r_{\rm n})=0 $, this condition becomes
\begin{align}
\label{}
2g^2 |n_0|(r_{\rm n}-m_0)=1 \,, \qquad n_0<0 \,. 
\end{align} 
Close to the nut surface $r=r_{\rm n}$, we have 
\begin{align}
\label{}
\check f^{-1}=\frac{-2n_0(r-r_{\rm n})}{r_{\rm n}(r_{\rm n}-2m_0)}+O((r-r_{\rm n})^2)\,, \qquad 
\check \Delta =\frac{(r-r_{\rm n})^2}{r_{\rm n}(r_{\rm n}-2m_0)}+O((r-r_{\rm n})^3)\,. 
\end{align}
In particular, both the fibre and the two-sphere shrink at $r=r_{\rm n}$, so that the fixed-point set of
$\partial/\partial t$ is zero-dimensional.  In terms of 
\begin{align}
\label{}
\rho^2=8|n_0| (r-r_{\rm n}) \,, \qquad \psi=\frac{t}{2n_0}\,, 
\end{align}
the metric around $\rho\simeq 0$ is approximated by (\ref{nutmetric}). 
The periodic identification required to remove the
Dirac-Misner strings implies $\Delta\psi=\Delta t/(2|n_0|)=4\pi/k$ ($k\in \mathbb N$). 
Thus, a smooth nut requires $k=1$.

\section{Conclusion}
\label{sec:conclusion}

In this paper, we have presented and analyzed a family of exact Taub-NUT-AdS solutions with scalar hair in four-dimensional
${\cal N}=2$ gauged supergravity.  These non-BPS solutions belong to an Einstein-scalar sector with
all gauge fields switched off, and their matter content is supplied entirely by a nontrivial scalar field and its self-interaction potential.
We have also made explicit their relation to a cohomogeneity-one scaling limit of the more general Einstein-scalar family of
\cite{Anabalon:2012ta}, thereby complementing the independent construction employed in this work.

Although the spacetime is plagued by a causal pathology and geodesic incompleteness as in the ordinary Taub-NUT-AdS metric,
the rich interplay between the NUT charge and scalar hair leads to several distinctive features.
In particular, the NUT deformation can give rise to a regular Killing horizon in the plus branch, whose static limit is horizonless.
We have determined the necessary and sufficient conditions for the existence of a horizon in the exterior region and shown that
at most one nondegenerate horizon can occur.  
The NUT charge and the slowly decaying scalar modes also leave distinct imprints on the asymptotic structure.
The former deforms the boundary geometry, while the latter backreact on the metric and contribute to the conserved mass.
By specifying mixed boundary conditions for the scalar field and treating the NUT parameter as fixed boundary data, we have obtained a finite and integrable mass variation and formulated a restricted thermodynamic prescription on the resulting phase space. 
The resulting restricted first law involves a thermodynamic Misner charge distinct from the geometric NUT charge. 
These results provide an analytically tractable setting in which to investigate the interplay between scalar hair, nontrivial
boundary geometry, and conserved quantities. Their possible holographic interpretation deserves further
study, with due attention to the causal properties of the Lorentzian boundary.

The Euclidean sector provides a complementary perspective.
In section~\ref{sec:Euc}, we have established the ambi-K\"ahler structure
of the generic Euclidean continuations and identified an
exceptional NUT-charged family at $\gamma=0$.
Restricting attention to the cases in which the higher-dimensional supergravity embedding is possible ($\gamma=0, \pm 1/2$), 
we have examined the existence and regularity conditions for bolts and nuts, including the 
role of the periodic identifications in distinguishing smooth nuts from orbifold quotients.
These results provide a starting point for assessing the semiclassical significance of the Euclidean solutions through
their renormalized on-shell actions and possible contributions as gravitational instantons.
The search for supersymmetric extensions relevant to localization is another direction worth pursuing.

Several questions remain open. 
An important next step is to investigate linear stability under
coupled metric and scalar perturbations. 
In particular, it would be interesting to determine how the 
NUT deformation affects the perturbation spectrum and whether
the two branches exhibit qualitatively different stability properties.
The cohomogeneity-two parent family also offers a natural 
starting point for studying less symmetric configurations within the same supergravity model, including their horizon
structure and global regularity.

Relaxing the Einstein-scalar truncation by restoring electromagnetic
and axionic fields would allow us to examine how these additional
degrees of freedom modify the horizon structure and the conditions for supersymmetry.
Constructing analogous solutions in higher-dimensional gauged supergravities would help distinguish features specific to
four dimensions from those that persist more generally.
Finally, it would be valuable to extend the analysis to a broader
class of scalar boundary conditions, which may require enlarging
the solution space beyond the exact family considered here.
Such an extension could also clarify the definition of conserved
charges and the thermodynamic formulation when variations of
the NUT parameter are allowed.

%
%
%
%
%
%
%

\acknowledgments
This work is partially supported by JSPS KAKENHI Grant Number 25K07309.



\end{document}